\documentclass[prd,aps,twocolumn,a4paper,showkeys,nofootinbib]{revtex4-1}

\usepackage{amsmath}
\usepackage{amsfonts}
\usepackage{amssymb}	
\usepackage{graphicx}
\usepackage{amsbsy}
\usepackage{color}
\usepackage{commath}
\usepackage{hyperref}

\allowdisplaybreaks

\newcommand{\TEOB}[1]{\texttt{TEOBResumS{#1}}}

\begin{document}

\title{Actively Learning Numerical Relativity}

\author{Tomas \surname{Andrade}${}^{1}$}
\author{Rossella \surname{Gamba}${}^{2,3,4}$}
\author{Juan \surname{Trenado}${}^{1}$}

\affiliation{${}^1$ Departament de F{\'\i}sica Qu\`antica i Astrof\'{\i}sica, Institut de
Ci\`encies del Cosmos, Universitat de
Barcelona, Mart\'{\i} i Franqu\`es 1, E-08028 Barcelona, Spain}

\affiliation{${}^2$Theoretisch-Physikalisches Institut, Friedrich-Schiller-Universit{\"a}t 
Jena, 07743, Jena, Germany}  

\affiliation{${}^{3}$ Institute for Gravitation \& the Cosmos, The Pennsylvania State University, University Park PA 16802, USA}

\affiliation{${}^{4}$ Department of Physics, University of California, Berkeley, CA 94720, USA}

\begin{abstract}

Data analysis of gravitational waves detected by the LIGO-Virgo-Kagra collaboration and future observatories relies on precise modelling of the sources. In order to build, calibrate and validate current models, we resort to expensive simulations in Numerical Relativity (NR), the fully-fledged simulation of Einstein's Equations. Since simulation costs and the dimensionality of parameter space are prohibitive to perform a dense coverage, approximate models interpolate among the available simulation data. 
We put forward the technique of Gaussian Process Active Learning (GPAL), an adaptive, data-driven protocol, for parameter space exploration and training of gravitational wave approximants. 
We evaluate this proposal by studying a computationally inexpensive scenario, in which we calibrate the approximant \TEOB{} using the NR-informed model as a proxy for NR. 
In this case study, we find that GPAL reduces the computational cost of training by a factor of 4 with respect to uniform or randomly distributed simulations. Moreover, we consider a parallel implementation which reduces computational time, and hybrid strategies which improve pre-calibrated models. 
The Gaussian Process regression employed in this approach naturally endows the algorithm with notion of model uncertainty. We comment on the implications of this feature for data analysis. 

\end{abstract}

\date{\today}

\maketitle

\section{Introduction}

The detection of gravitational waves by the Ligo-Virgo-Kagra collaboration in 2015 opened a new scientific era \cite{LIGOScientific:2016aoc}. This emerging multidisciplinary endeavour combines detectors of exquisite precision and sophisticated data analysis techniques \cite{Jaranowski:2005hz}. 
Due to the large amount of experimental noise, the most sensitive searches involve matched filtering, in which one compares the data stream to a large template bank of candidate signals \cite{LIGOScientific:2011hqo}. Moreover, the characterization of the physical sources is conducted via parameter estimation, a Bayesian inference technique which also compares the detected signal with theoretical templates \cite{LIGOScientific:2013yzb}. 
Both aspects highlight the crucial role of the precise mathematical modelling of gravitational waves. 

For concreteness, henceforth we restrict ourselves to the case of black hole binaries (BBH) described by General Relativity. These are by far the most numerous events detected to date, with 90 BBH events recorded in the observational campaign O3 \cite{LIGOScientific:2021djp}, number which will significantly increase with the data currently being collected by O4 \cite{KAGRA:2013rdx}.

The fully fledged simulation of General Relativity is termed Numerical Relativity (NR), and it is by now a mature field, at least in the absence of matter sources. The first successful simulations of a black hole binary in 2005 \cite{Pretorius:2005gq, Campanelli:2005dd, Baker:2005vv}, paved the way for many developments which have materialized in a series of open source codes \cite{spec, ETK, Daszuta:2021ecf, Andrade:2021rbd} which allow researchers to carry out these simulations routinely. 
The complexity of the problem at hand is that of exploring a large parameter space  -- 10 dimensions of intrinsic parameters --  with highly expensive simulations -- the average cost is of the order of $6 \times 10^3 $ CPU hours (approximately 2 days with 100 processes running in parallel) \cite{varma_thesis}. 
The most common strategy to alleviate this technical challenge is to resort to {\it approximants}, i.e. analytic or semi-analytic\footnote{Here by semi-analytic we mean models that require solving ODEs. While requiring some numerical computation, the cost of this is several orders of magnitude smaller than that of NR which requires solving PDEs.} approximate models which in turn require calibration by a reduced set of NR simulations. Once properly trained, these models are fast to evaluate and provide a good basis to build template banks. 
We can distinguish 3 main families of approximants: i) phenomenological models 
\cite{Ajith:2007qp, Ajith:2007kx, Ajith:2009bn, Santamaria:2010yb, Husa:2015iqa, Khan:2015jqa, Pratten:2020fqn, Estelles:2020osj, Hamilton:2021pkf, Estelles:2021gvs,  London:2017bcn, Garcia-Quiros:2020qpx, Khan:2019kot, Hannam:2013oca, Schmidt:2014iyl, Khan:2018fmp, Pratten:2020ceb}; 
ii) Effecive One Body \cite{Buonanno:1998gg,Buonanno:2000ef,Damour:2000we,Damour:2001tu,Damour:2008gu, Damour:2008qf,Barausse:2009xi,Damour:2009sm,Damour:2009wj,Damour:2016gwp,Vines:2017hyw,Damour:2017zjx}, implemented via two sub-families, SEOBNRv* \cite{Bohe:2016gbl,Babak:2016tgq,Cotesta:2018fcv,Ossokine:2020kjp, vandeMeent:2023ols, Ramos-Buades:2021adz, Pompili:2023tna, Khalil:2023kep, Ramos-Buades:2023ehm} and \TEOB{} \cite{Damour:2014yha, Nagar:2017jdw,Nagar:2018zoe,Akcay:2020qrj,Gamba:2021ydi,Nagar:2020pcj, Chiaramello:2020ehz, Nagar:2020xsk, Albanesi:2021rby, Nagar:2021gss, Nagar:2021xnh,Nagar:2023zxh}, and iii) NR Surrogates \cite{Blackman:2014maa, Blackman:2017dfb, Varma:2018mmi, Varma:2019csw, Williams:2019vub}. 
The details of the training strategy depend on the family of approximant under consideration. 
The typical procedure is to construct the training set from pre-existing NR databases, which are built relying on domain expert knowledge, e.g. range of parameters believed to be astrophysically relevant, technical feasibility of carrying out the simulations for certain parameters, etc. 
The test set is also usually chosen from existing NR simulation data, but sometimes other approximants are used for validation in certain regimes. 
The main NR open databases are the 
SXS~\cite{Mroue:2013xna,Chu:2015kft,Boyle:2019kee}, 
RIT~\cite{Healy:2017psd, Healy:2019jyf, Healy:2022wdn}, 
CoRe~\cite{Dietrich:2018phi, Gonzalez:2022mgo}, 
SACRA-MPI \cite{Kiuchi:2017pte,Kiuchi:2019kzt}, Cardiff~\cite{Hamilton:2023qkv} 
and MAYA~\cite{Ferguson:2023vta} catalogues.

%

Our goal here is to describe a systematic protocol to chose a series of NR simulations for GW model calibration in order to make them as informative as possible, thus reducing computational costs and alleviating human bias. 
The key observation is that the problem of exploring parameter space with expensive simulations is precisely the goal of Experimental Design \cite{10.1214/aoms/1177728069, Atkinson2007},
a field of statistics which is concerned with the determination of a series of ``experiments'' (here NR simulations) which can best inform and validate a given model (here the waveform approximants). 
As we will describe, can easily port to the problem at hand the methodology of Gaussian Process Active Learning (GPAL) \cite{CHANG2021101360}, developed in the field of Cognitive Science, building upon the literature of Bayesian Optimization Experimental Design \cite{10.1214/ss/1177009939} and Active Learning \cite{cohn1996active, agnihotri2020exploring} in Machine Learning\footnote{Note that this is unrelated to the Bayesian Inference approach employed in parameter estimation for gravitational waves.}.
The main feature of this approach is to take a set of existing simulations as Bayesian priors, and suggest the next experiment as a posterior maximizing a certain acquisition function, which captures the degree of new information gained by performing this new experiment. 
In this context, the natural acquisition function is the variance of the Gaussian Process (GP) regression \cite{krige1951statistical, 10.7551/mitpress/3206.001.0001, wang2022intuitive} used to fit the data, which intuitively captures the idea of maximizing information at the regions of parameter space where the variance is highest. 
This and related {\it adaptive} methods have proven useful to increase model accuracy and reduce experimental costs in a variety of contexts
\cite{Cavagnaro2011, Cavagnaro2013, 10.1167/16.6.15, 10.1115/1.4057002, foster2020variational, foster2021deep} 
%
as we also observe in our problem of interest. 

As an interesting byproduct of this approach, the GP variance can be used to construct a measure of model uncertainty. We shall leverage this feature to endow the approximant with a precise notion of local (input specific) waveform uncertainty and error estimates in the time domain. 
This has been previously noted in \cite{Williams:2019vub}, which modelled the strain time series as GP and used the variance to provide error estimates in the time domain. 
The relevance of this model uncertainty for parameter estimation has been put forward in \cite{Moore:2014pda, Moore:2015sza}, and recently considered in \cite{Breschi:2022xnc, Read:2023hkv}.

As a concrete example, we will calibrate the EOB model \TEOB{}, taking as a proxy for NR simulations a previously calibrated version of \TEOB{} itself. This will serve as a proof of concept revealing the potential and challenges of the approach, while reducing the computational costs to a minimum. 
Based on this case study, we estimate that the active approach to build a training set of NR simulations can reduce the computational cost by a factor of 4x with respect to uniform or randomly chosen NR simulations. We also discuss a parallel implementation of the model which help reduce computational time, and hybrid strategies which could be used to improve previously calibrated models. 

This paper is organized as follows. 
In Sec. \ref{sec:model} we review the waveform model \TEOB{}, paying special attention to the internal parameters which require NR information. 
We lay out the methodology in Sec. \ref{sec:method}, focusing on the general use of the GPAL algorithm for time series, and its applications to GW modelling. 
We present our results in Sec. \ref{sec:results}, showcasing the calibration improvements and introducing the notion of waveform uncertainty. 
We conclude in Sec. \ref{sec:conclusions}, summarizing the advantages of our approach. 
We discuss some limitations of the scope of our case study, and some potential limitations of the algorithm itself in Sec. \ref{sec:discussion}. 
Finally, we put forward some concrete open problems suggested by the current study in Sec. \ref{sec:outlook}. 
We collect some relevant details of GP regression and additional numerical results in the appendices.

\section{Waveform model}
\label{sec:model}

\begin{figure*}[] 
\centering
\includegraphics[width=1\textwidth]{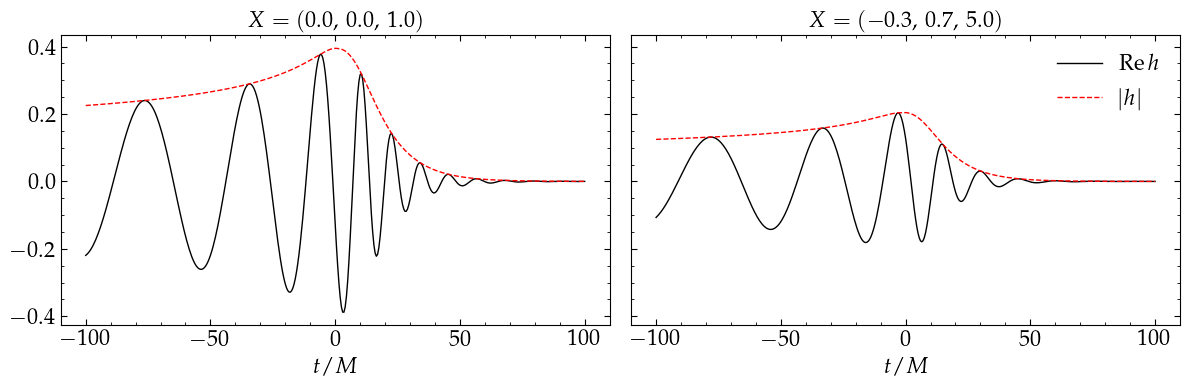}
\caption{Selected waveforms with initial data $X = (\chi_1, \chi_2, q)$ for a spinless binary with $q = 1$ (left) and a spinning binary with $q = 5$ (right).}  
\label{fig:wf_ex}
\end{figure*}

Our main tool in this study will be the EOB model \TEOB{}, a state-of-the-art approximant for spinning compact binaries (BBHs, BNSs or BHNSs) coalescing along generic orbits. 
The EOB approach is based on the mapping of the dynamics of the general relativistic two-body problem into the motion of a test mass in an effective background metric. By further resumming high order post-Newtonian results, and augmenting them via NR information, \TEOB{} can robustly and efficiently generate waveforms throughout the entire inspiral-merger-postmerger phases of a coalescence. 
There are several versions of this model publicly available. We will base our work on $\tt{v3.0.0}$ of \TEOB{-GIOTTO} \cite{Nagar:2018zoe, Riemenschneider:2021ppj, Gamba:2021ydi}. 
Although this version of the model is not the most developed for the case of quasi-circular (QC) orbits, all results obtained below can be straightforwardly applied to the more recent versions of the model. From now on, we shall refer to this model simply as \TEOB{}. 
Within the QC case, we shall set the precession to zero, which implies that the black hole spins are aligned with the orbital angular momentum. Therefore, the main physical input parameters for \TEOB{} are the mass ratio $q = m_1/m_2$ (where $m_1$, $m_2$ are the black hole masses), and the two dimensionless spin parameters $\chi_1$, $\chi_2$, which take values between $-1$ and $1$, where the sign accounts for the orientation of the spin with respect to the orbital angular momentum. 
The code can generate waveforms for any value of mass ratio $q\geq 1$ and spins $|\chi_{1,2}| \leq 1$, and has been validated against NR simulations up to values of $q=128$~\cite{Nagar:2022icd}. 
In addition, we can specify an initial frequency $f_{\rm 0}$ -- or alternatively an initial radial distance $r_{0}$ -- which controls the starting point of the dynamics. As such, this quantity is not an intrinsic parameter of the source and does not stand on the same footing as the mass ratio or spins. We set the starting point to $r_{\rm 0} = 14 M$ in geometric units without loss of generality, corresponding to an initial frequency of approximately $0.006$.
For given values of the initial data $(\chi_1, \chi_2, q)$, the code outputs a set of complex time series, denoted as modes, labelled by $(\ell, m)$ referring to the Spherical Harmonic decomposition. Here we will focus on the leading mode $(\ell, m) = (2,2)$, and omit the mode indices henceforth. We show some selected examples of the waveforms in Fig \ref{fig:wf_ex}. 

As mentioned above, the model depends on some analytically known parameters (e.g. the EOB potentials expressed as resummed Post-Newtonian expansions), and some free parameters which are fixed by calibrating (or informing) the model by comparing the resulting waveforms to a set of NR simulations taken as the ground truth. 
More precisely, denoting the initial data collectively as $X$, NR produces the ground truth signals 
\begin{equation}
\label{eq:h_gt_NR}
    h_{gt}(X, t) = h_{\rm NR}(X; t)
\end{equation}
Denoting the internal model parameters which require calibration as $c_i(X)$, the EOB model produces an approximate waveform $\tilde h(X, t)$, 
\begin{equation}
\label{eq:h_tilde}
    \tilde h(X, t) = h_{\rm EOB}(c_i(X); t)
\end{equation}
The internal parameters of the model are given by smooth functions of the mass ratio and spins, which typically take the from of ratios of polynomials. In this work we will focus on six internal parameters $c_i = \{A_{\rm mrg}, \omega_{\rm mrg}, A_{\rm NQC}, \omega_{\rm NQC}, \dot A_{\rm NQC}, \dot \omega_{\rm NQC} \}$ (note that we suppress mode indices since we focus on the 22 mode). The parameters $A_{\rm mrg}, \omega_{\rm mrg}$ correspond to the amplitude and frequency at merger, while the remaining parameters ensure that the waveforms are smooth near the peaks, see e.g. Ref.~\cite{Nagar:2020pcj, Riemenschneider:2021ppj} for a precise definition. 
The dependence of the model parameters with the initial data  is given by smooth, slowly varying functions, which can be written as products of ratios of polynomials that contain arbitrary coefficients to be fixed upon informing the model with NR, see Appendix \ref{app:eob}. 

Calibrating or informing the model amounts to 
\begin{itemize}
    \item Choosing a set of points $\{X\}^{{train}}$, the {\it training set}, representative of some region of parameter space the phenomenology of which we want to capture

    \item Extracting the values of the internal parameters $c_i(\{X\}^{{train}})$

    \item Providing global fits $c_i(X)$ which interpolate among the $c_i(\{X\}^{{train}})$
    
\end{itemize}
\noindent while minimizing one or more target metrics which attempt to capture how similar $\tilde h(X, t)$ and $h_{gt}(X, t)$ are  evaluated over some test set $\{X\}^{{\rm test}}$. 
A common choice of such metrics is the unfaithfulness or mismatch (see Appendix \ref{app:metrics}). 
In addition to the unfaithfulness, \TEOB{} aims to minimize the phase and amplitude differences at merger \cite{Nagar:2023zxh}, since it has proven useful to incrementally improve different aspects of the model. The version of the model used here has been informed with about 130 simulations, see \cite{Nagar:2020pcj} for details.


\section{Methodology}
\label{sec:method}

As stated above, calibration of a given model requires carrying out NR simulations in a judiciously chosen training set $\{X\}^{{\rm train}}$. 
Here we consider GPAL, an implementation of Bayesian Optimal Experimental Design which deploys Active Learning in combination with Gaussian Process regression. In particular, this strategy selects the training set $\{X\}^{{\rm train}}$ iteratively, maximizing the variance of the Gaussian Process fits performed on the parameter space after each iteration. 
This captures the intuitive idea that every new simulation is the one that provides more ``information'' about the system. 

\subsection{GPAL for time series}

Let us briefly describe the GPAL algorithm and the relevant modifications we introduce for our study. 
For a given target $f(X)$ -- the function to be fitted or ``learned'' -- defined on a domain $D(X)$, the GPAL algorithm can be summarized as follows. As the initial step, give prior values $\tilde f(X)$ and an initial variance $\sigma(X)$. Then, iterate over
\begin{enumerate}
    \item Evaluate $f$ at the maximum of $\sigma(X)$, $X_0$

    \item Taking into account the newly extracted value $f(X_0)$, compute the posterior of $f_{{\rm post}}(X)$ as the Gaussian Process regression of the function on $D(X)$, with new variance $\sigma_{{\rm post}}(X)$. 

    \item Update $\tilde f(X) = f_{{\rm post}}(X)$ and $\sigma(X) = \sigma_{{\rm post}}(X)$
\end{enumerate}
Thus, at after each iteration, we obtain a training set $\{X\}^{{\rm train}}$, and approximate function $\tilde f(X)$ and a variance $\sigma(X)$. 
We have used GP regression with kernel (covariance function) Matern52, with $\sigma = 1$, implemented in the library GPy \cite{GPy}, see Appendix \ref{app:GP} for details. 

In the application at hand, we use GPAL to evaluate the internal model parameters $c_i(X)$ which play the role of the function $f(X)$ in the discussion above\footnote{We have done so by fitting each parameter individually as a function of $X$. Other ``multi-output'' choices are possible in which one can take into account  correlations among the various functions to be approximated  \cite{NIPS2007_66368270}.}
This has allowed us to deploy this strategy to the case of time series. Note that, as opposed to \cite{Moore:2014pda, Moore:2015sza, Williams:2019vub}, we are not describing the time series as GP, but as given by the fixed (time) functional forms specified by the model at hand from the onset.

The algorithm easily allows for other ways of selecting the evaluation points, or {\it acquisition functions}, instead of maximising $\sigma(X)$. One that appears particularly useful in the case where there is high variability in a particular region of parameter space is the norm of the gradient of $f$ (or even $\tilde f$, to reduce evaluation costs). We leave this investigation for future work. 

\subsection{EOB as proxy for NR}

In order to carry out an in-depth study of the calibration process with minimal computational cost, we shall take use EOB as a proxy for NR. 
More specifically, we take the NR informed approximant \TEOB{} as the ground truth, i.e. 
\begin{equation}
\label{eq:h_gt}
    h_{gt}(X, t) = h_{\rm EOB}(c_i(X); t),
\end{equation}
and use it to calibrate an un-informed version of the model 
\begin{equation}
\label{eq:h_tilde}
    \tilde h(X, t) = h_{\rm EOB}(\tilde c_i(X); t)
\end{equation}
\noindent Note that the ground truth and approximate models in \eqref{eq:h_gt} and \eqref{eq:h_tilde} only differ by the values for the internal parameters, which are tuned $c_i(X)$, and detuned $\tilde c_i(X)$. As mentioned above, we detune the internal parameters corresponding to $c_i = \{A_{\rm mrg}, \omega_{\rm mrg}, A_{\rm NQC}, \omega_{\rm NQC}, \dot A_{\rm NQC}, \dot \omega_{\rm NQC} \}$. 
Since these affect more strongly the late inspiral and ring-down of the waveforms, we restrict our study to a time interval $(-100 M, 100 M)$ centered at the peak of the absolute value of each signal.

We extract the internal parameters for the EOB approximant simply by calling the output of the code. This eliminates uncertainties in the extraction, which are present in a realistic case\footnote{For example, in order to extract the amplitude of the waveform at merger, we would have to evaluate the value of the waveform at the amplitude peak. It turns out that there are minor differences between the direct output value and the extracted value, which introduce extra difficulties in the calibration.}. 
We will discuss some of the limitations of our approach in the Sec. \ref{sec:discussion}.

\subsection{Calibration strategies}
\label{sec:calibration}

To bench mark the GPAL strategy, we consider two alternative approaches consisting of uniformly spaced, and random grids in parameter space. We will compare the results provided by these different choices using the same number of training points in each approach.  
We choose the input parameters in the range $-0.85 \leq \chi_{1,2} \leq 0.85$, $1 \leq q \leq q_{max}$, and study the algorithm behaves as we vary $2 \leq q_{max} \leq 6$. We discretize the space of parameters by choosing a uniform grid with $64^3$ points in the directions $\chi_1$, $\chi_2$, $q$.  

We initialize the active algorithm by choosing flat priors for all internal parameters $c_i(X) = 1$ and their variance $\sigma(X) = 1$, and picking the first iteration to be a random point. We observe that the subsequent iterations tend to lie at the boundaries of the computational domain. After a certain number of iterations, internal points get selected\footnote{Since we take $\sigma(X)$ to be the acquisition function and initialize the algorithm with flat priors, the selected training points and fit variance are independent of the functions being fitted, see Appendix \ref{app:GP}. In particular, this means that the variance is the same for each internal parameter $c_i$.}. 
We choose the uniform training grids by specifying points on the boundary of the domain and sub-sampling each direction by the same number of points, e.g. for a 1d grid of 64 points, a sub-sampling of 32 yields 3 equally spaced nodes in each direction, so that the number of training points is $3^3$. 
The points of the random grid are selected without restrictions, so they may are may not cover the regions near the edges of the domain. 
The test set for each numerical experiment is chosen to be the reciprocal lattice of the corresponding homogeneous training grid, including boundary points equidistant from it. 

\begin{figure}[] 
\centering
\includegraphics[width=0.5\textwidth]{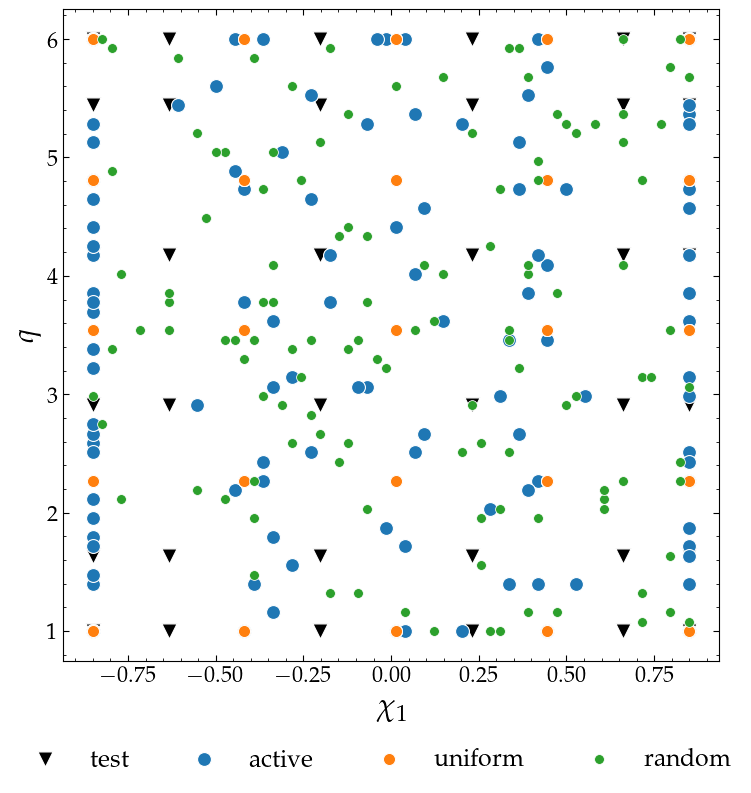}
\caption{Training and test sets for $N = 125$, projected on the 2D slce $(\chi_1, q)$.} 
\label{fig:training_test_coords}
\end{figure}

In this setup, we monitor how the performance and uncertainty of the algorithm changes as we vary the number of training points. For our numerical experiments we take the sub-sampling of the uniform grid, number of training points, and size of test set as given in Table \ref{tab:N}.
We depict the resulting training sets for $N = 125$ in Fig. \ref{fig:training_test_coords}.

\begin{table}[h]
\begin{tabular}{lllllllll}

\hline
\hline
$1d$ sub-sample &  32 & 22 & 16 & 13 & 11 & 9 & 8  \\
\hline
$N_{{\rm train}}$  &  27 & 64  & 125 & 216 & 343 & 512 & 729 &  \\
\hline
$N_{{\rm test}} $ &  64 & 125  & 216 & 343 & 512 & 729 & 1000 & \\
\hline
\end{tabular}
\caption{Sub-sample step size in each direction of the grid, number of training points, and number of test points for each epoch of our numerical experiments.}
\label{tab:N}
\end{table}

\subsection{Model assessment}
\label{sec:model_assessment}

At the end of each calibration round, we compute various performance and uncertainty metrics, as we now describe. 
For a given point in the test set $X$, we extract the ground truth model parameters $c_i(X)$ and the fitted values $\tilde c_i(X)$, and evaluate the ground truth and approximate signals using \eqref{eq:h_gt}, \eqref{eq:h_tilde}. 
We then align -- adjust the relative time and phases -- the so obtained signals with respect to one another using the Python package pyCBC.

The main performance metric we consider is the unfaithfulness between the ground truth and approximate signals 
\begin{equation}
\label{eq:metric_unfaith}
    \overline{{\cal F}}(X) = \overline{{\cal F}}(h_{gt}(X), \tilde h(X))
\end{equation}
This is a standard quantity to characterize the notion of ``distance'' between two waveforms in the context of GW data analysis. In addition, we consider the amplitude $\delta A(X)$ and phase differences $\delta \phi(X)$ at merger between the ground truth and approximate signals at a given point in parameter space $X$, similarly to \eqref{eq:metric_unfaith}. 
As argued in \cite{Nagar:2023zxh}, these merger differences provide useful guidance in waveform modelling. See  Appendix \ref{app:metrics} for the precise definition of these quantities. 
These metrics are chosen such that a model with perfect performance has $\overline{{\cal F}}(X) = \delta A (X) = \delta \phi(X) = 0$ across the entire test set. 

   
The variance of the Gaussian fits, $\sigma(X)$ endows the algorithm with a natural measure of uncertainty at each point in parameter space. 
In order to make this more easily interpretable in the context of GW data analysis, we introduce the notion of {\it waveform uncertainty} $u(X)$ by considering the unfaithfulness between the approximate signal at zero variance and the most uncertain signal given the variance of all internal parameters, $\tilde h^{(\alpha_0)}(X)$, see Appendix \ref{app:metrics} for a precise definition. The waveform uncertainty can be expressed as
\begin{equation}
\label{eq:u}
    u(X) = \overline{{\cal F}}\left(\tilde h(X), \, \tilde h^{(\alpha_0)}(X) \right)
\end{equation}
Finally we consider the fit accuracy, defined as the Euclidean norm of the difference between the ground truth values of the parameters and the fitted ones, which we denote as $\delta c(X)$. 

We will typically be interested in the average values of these metrics on the test set, which we denote as 
\begin{equation}
\label{eq:test_set_mean}
    \langle M \rangle = \frac{1}{N_{{\rm test}}} \sum_{i= 1}^{N_{{\rm test}}} M(X_i) 
\end{equation}

\noindent with $M$ being any of the aforementioned metrics, and $i$ running over the $N_{{\rm test}}$ points in the test set $\{X\}^{{\rm test}}$. 

\section{Results}
\label{sec:results}

\subsection{Performance and calibration cost}
\label{sec:performance}

We compare the three training strategies, active, uniform and random, by carrying out each training with a given number of points, and evaluating the performance and uncertainty metrics. 

\begin{figure*}[] 
\centering
\includegraphics[width=0.8\textwidth]{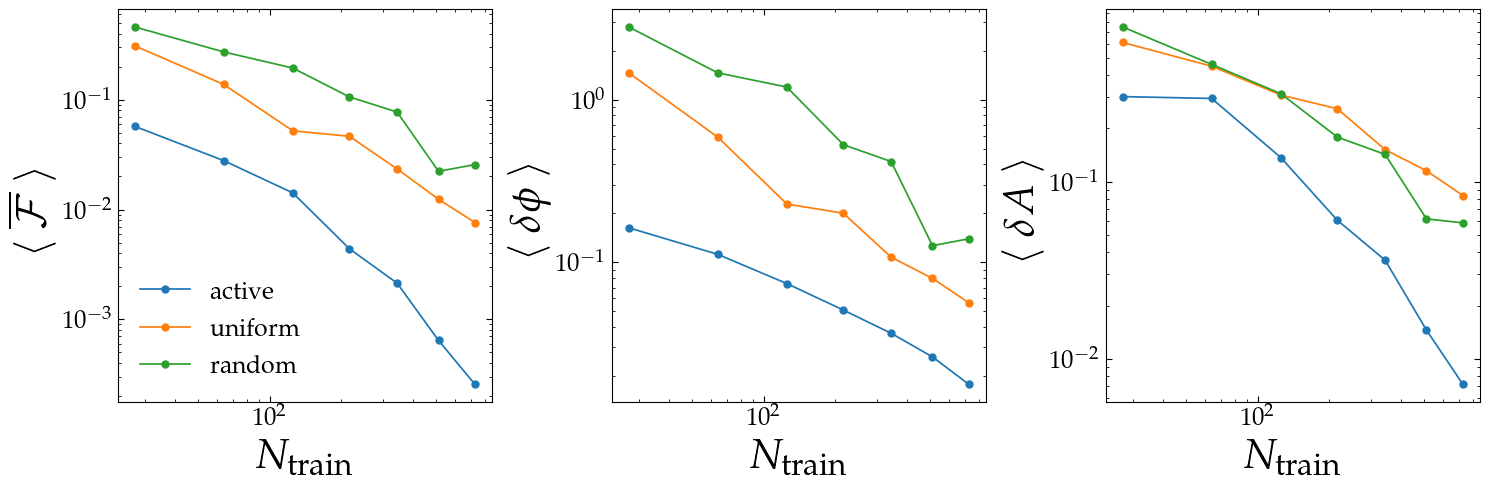}
\includegraphics[width=0.8\textwidth]{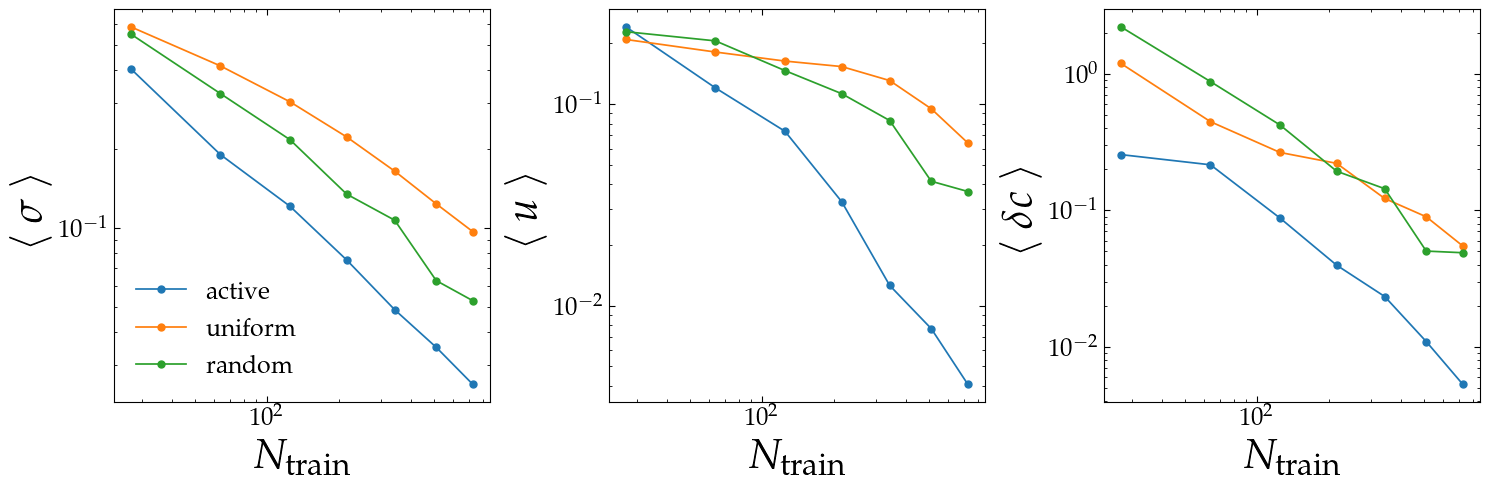}
\caption{Dependence of the average performance (top) and uncertainty metrics (bottom) with number of training points. All plots are in log-log scale.} 
\label{fig:metrics_vs_Ntrain}
\end{figure*}

We begin discussing the average values of the metrics over the test set, constructed as explained in Sec. \ref{sec:model_assessment}. In Fig. \ref{fig:metrics_vs_Ntrain} we show the dependence of all metrics with the number of training points for $q_{{\rm max}} = 6$. We clearly appreciate that the active training strategy outperforms the uniform and random strategies. Interestingly, we observe an approximate power-law dependence with $N_{{\rm train}}$ in most of these quantities, with the exception of those associated to random training. We expect that averaging over realizations would bring the dependence of these quantities closer to power-laws. 
We obtain qualitatively similar results for smaller values of $q_{max}$, although the performance difference between the active and other training protocols increases with $q_{max}$. 

\begin{figure}[] 
\centering
\includegraphics[width=0.4\textwidth]{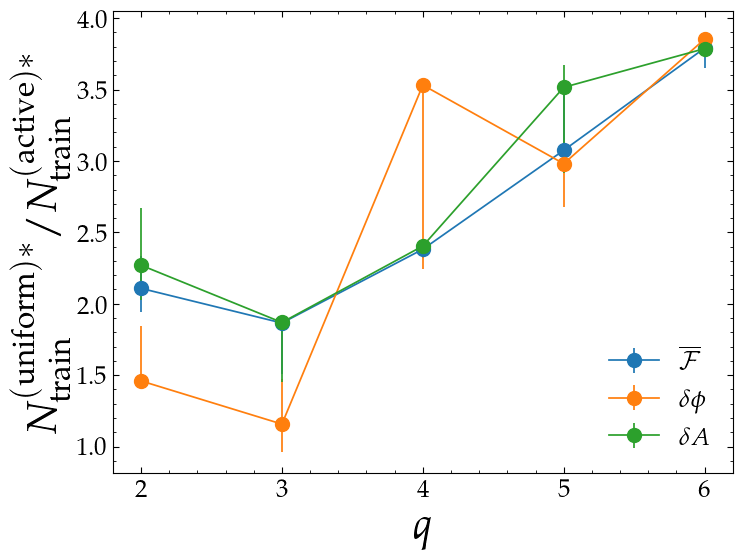}
\caption{Average performance gain for all performance metrics, as a function of $q_{max}$. We obtain the error bars for each data point by varying the base-line values by $\pm 30 \%$, computing all intercepts, and keeping the highest and lowest values.} 
\label{fig:performance}
\end{figure}

Taking as a reference target value $\overline{{\cal F}} = 0.01$, we interpolate the performance curves and extract the number of training points $N_{{\rm train}}^{({\rm type}) *}$ required to achieve this performance target, with ${\rm type} = {\rm active}, {\rm uniform}, {\rm random}$
We can then estimate the cost reduction provided by the active strategy with respect to the uniform/random ones, by computing the number of training points required to achieve the corresponding targets, and taking their ratio\footnote{Note that this estimate assumes that the cost of all simulations is the same, which is not the case in a realistic scenario, see Sec. \ref{sec:discussion}}. 
Carrying out the same analysis with the amplitude and phase differences at merger with reference values $\delta A = 0.1$, $\delta \phi = 0.06$ yields the results in Fig. \ref{fig:performance}. We observe a maximum cost reduction of around 4x for all the metrics considered. 

We have also studied the dependence of the various metrics as a function of the location in parameter space. We observe that performance and uncertainty metrics deteriorate significantly near the edges of the domain, see Appendix \ref{app:local}. 

We consider extra alternative test and training sets in Appendix \ref{app:more}. In particular, we monitor the effect of altering the test set by removing boundary points and making it fully random. 
Moreover, we consider a training set given by Chebyshev nodes along all directions, allowing a more dense coverage of the regions near the boundary. 
In all cases we see robust results for the supremacy of the active training, with a cost reduction of roughly 4x in most cases. 

\subsection{Parallelization and calibration time}
\label{sec:parallel}

In principle, the GPAL strategy as described above should be followed serially. However, it is possible to parallelize it by partitioning the computational domain to obtain the training points, and performing a global fit at the end. 
For concreteness, we partition the domain in $N_q = 17$ two-dimensional slices of constant $q$, corresponding to approximately $\{q^i\} = 1 + 5/16 i$, with $i = 0 .. 16$, explore each slice {\it in parallel} with $N^{(2d)}_{train} = 6, 8, 10, 12, 14, 16, 18$ simulations, and finally perform a global GP regression in the 3D domain. 
We then study the performance of the algorithm as we vary the total number of simulations $N^{\parallel}_{train} = N_q N^{(2d)}_{train}$ keeping $N_q$ fixed. 
We show our results in Fig. \ref{fig:performance}. We observe that in order to achieve the same performance as the one with $N_{train} = 125$ we require about $N_{train} = 187$ (corresponding to $N^{(2d)}_{train} = 11$), i.e. around $50 \%$ more simulations to reach the same performance considering all metrics. 

%
Let us call the computation time per simulation $t_{{\rm sim}}$. 
In Sec. \ref{sec:performance} we found that the performance achieved with $N^{{\rm active}}_{{\rm train}} = 125$ is comparable to the one resulting from $N^{{\rm uniform}}_{{\rm train}} = 512$. Assuming that we can carry out $N^{(2d)}_{train} = 17$ simulations in parallel, the time it takes to calibrate the model with a uniform grid is $T_{{\rm uniform}} = 512/17 t_{{\rm sim}} \sim 30 t_{{\rm sim}}$. 
On the other hand, since we found that $N^{(2d)}_{train} = 11$, building the training set in the parallel implementation takes $T^{\parallel}_{active} = N^{(2d)}_{train} t_{{\rm sim}} =  11 t_{{\rm sim}}$, approximately one third of the time required by the uniform grid calibration. 

\begin{figure}[] 
\centering
\includegraphics[width=0.4\textwidth]{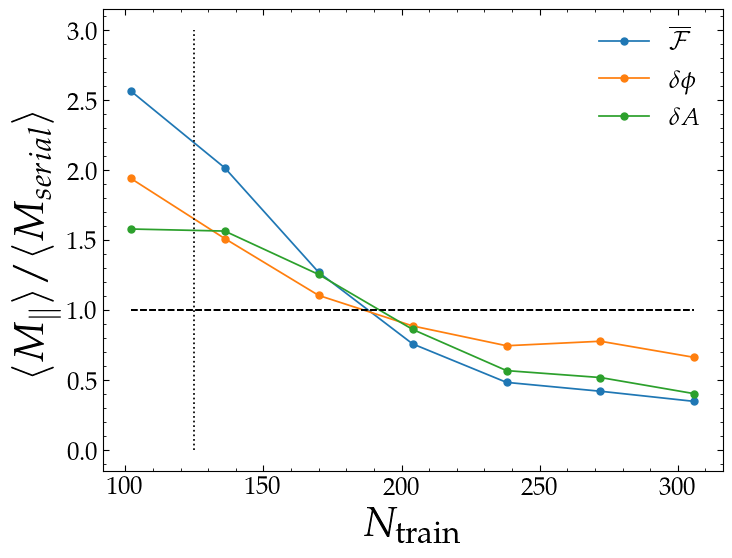}
\caption{Relative performance between the parallel and serial versions of our active training strategy, expressed as ratios of the performance metrics versus number of training points.} 
\label{fig:parallel}
\end{figure} 

\subsection{Hybrid strategy and model improvement}
\label{sec:hybrid}

The active approach can be also deployed to improve a model previously calibrated with a different strategy, as we will now illustrate. 

For concreteness, consider the case $q_{{\rm max}} = 6$ discussed in Sec. \ref{sec:performance}, focusing on the active and uniform grid cases. Instead of utilizing the active strategy from the onset, we can take as a starting point for the active algorithm a model calibrated with $N_{{\rm train}} = 125$ using a uniform grid, and then switching to active training. 

We see the outcome of this hybrid procedure in Fig \ref{fig:performance}, which also shows for comparison the purely active and purely uniform results. We observe that the hybrid training quickly catches up with the active one, and even outperforms it for large training sets.  
For the average unfaithfulness, we see an order of magnitude performance gain for $N_{{\rm train}} = 216$ of the hybrid protocol over the uniform one, and almost a two order of magnitude gain for $N_{{\rm train}} = 729$. The phase and amplitude differences at merger show a significant but slightly lower increase. 
%


\begin{figure*}[] 
\centering
\includegraphics[width=0.9\textwidth]{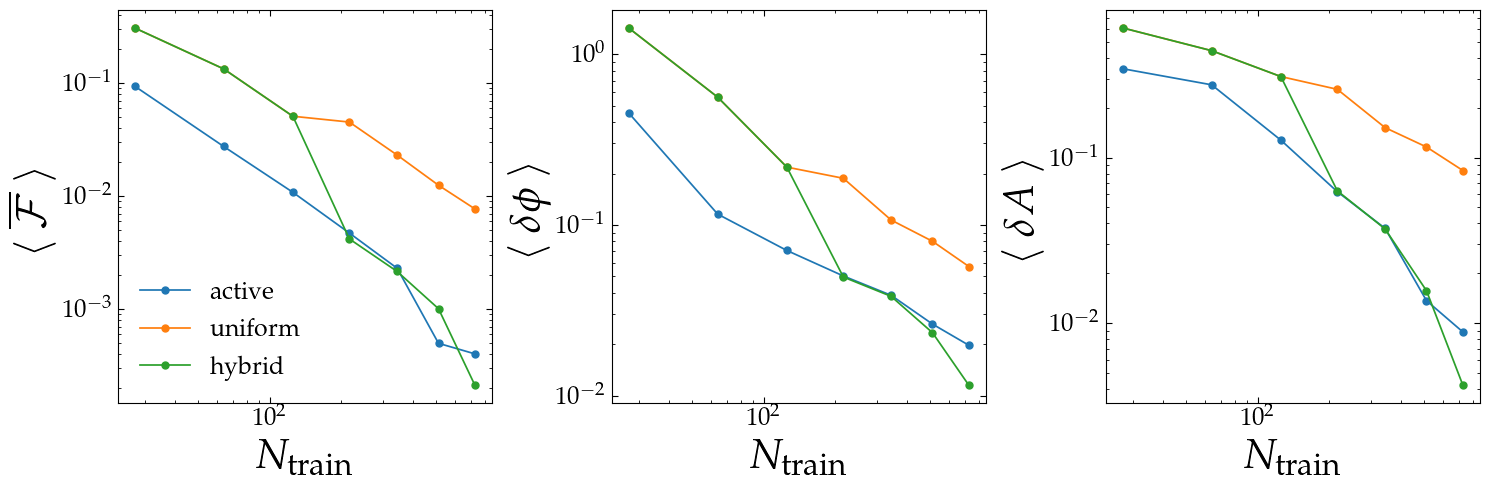}
\caption{Average performance metrics as a function of the number of training points for active, uniform and hybrid strategies. In the hybrid case, we initialize the training with the same prior as for uniform training with $N_{{\rm train}} = 125$ (hence the curves coincide up to that point), and switch to active training.} 
\label{fig:parallel}
\end{figure*}

\subsection{Waveform uncertainty}

As explained above, the variance of the Gaussian fits provides a natural way of capturing the model uncertainties across parameter space. Moreover, this can be translated into waveform uncertainties as explained in Sec. \ref{sec:model_assessment}. We have seen in Fig \ref{fig:metrics_vs_Ntrain} that the average waveform uncertainty as a function of training points decreases much faster with the active approach compared to uniform and random grids. In particular, we see that only for the active approach this uncertainty goes below the target unfaithfulness $\overline{{\cal F}} = 0.01$ in the ranges of parameters we have considered in this study.

In Appendix \ref{app:local} we show a histogram of the waveform uncertainties resulting from all training strategies at the maximum number of training points $N_{{\rm train}} = 729$. This again shows the better accuracy of the active approach, with only very few cases exceeding $\overline{{\cal F}} = 0.01$. 

\begin{figure*}[] 
\centering
\includegraphics[width=1\textwidth]{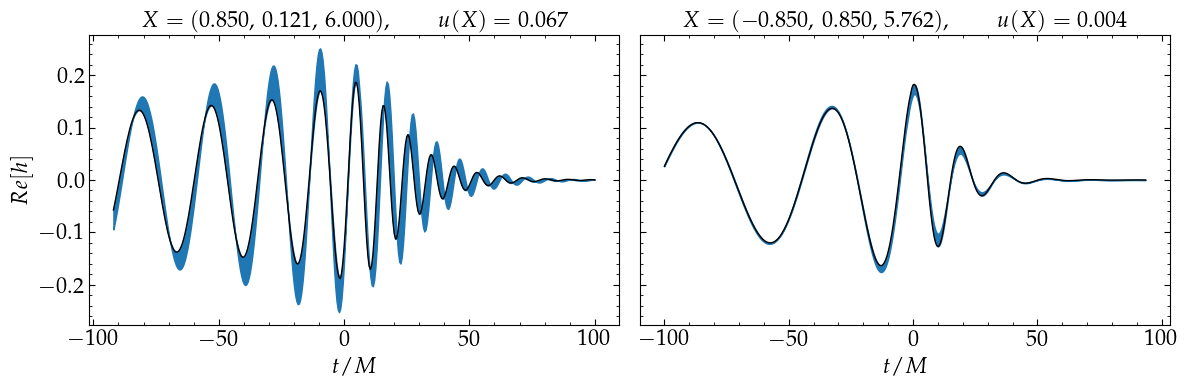}
\caption{Predicted waveform and the corresponding most uncertain waveform (as defined in Appendix \ref{app:metrics}) for the maximum (left) and average value (right) of the waveform uncertainty $u(X)$ in the case $q_{max} = 6$. We shade the region between both waveforms to ease visualization.} 
\label{fig:wf_unc_re}
\end{figure*}

Using this notion of uncertainty, we can provide error bars for the waveforms in the time domain, as shown in Fig. \ref{fig:wf_unc_re}. We leave for future work a careful consideration of the implications of this feature for data analysis. 

\section{Conclusions}
\label{sec:conclusions}

We propose the use Active Learning to systematically construct the training sets upon which waveform approximants are built.  
More precisely, we put forward the idea of adapting the technique GPAL to time series modelling. This involves performing a GP regression to fit the internal parameters from available NR data, and using the variance of the GP as the acquisition function to explore NR parameter space maximising the information gained from each simulation. 
Furthermore, we introduce the notion of waveform uncertainty, which is built out of the variance of the GP and provides a local, quantitative, and intuitive measure to endow the approximants with error estimates. 

This stands in contrast with the standard techniques used for model calibration, in which a fixed set of NR simulations is chosen based on domain knowledge (e.g. priors of astrophysical relevance, ease to carry out NR simulations), and global fits of the internal parameters are performed by postulating fixed functional forms. 
Moreover, with the notable exception of some surrogate models which also employ GP regression e.g. \cite{Williams:2019vub}, most of the available approximants typically only provide estimates of global inaccuracy (e.g. maximum unfaithfulness over a chosen test set) and do not provide local uncertainty estimates of the waveforms. 

Our strategy appears to be advantageous in the following ways: 
\begin{itemize}
    \item For a fixed performance target in various metrics, it requires fewer simulations compared to non-adaptive methods. We observe a cost reduction by a factor of up to 4x in the simple model considered here. Moreover, larger gains occur for larger parameter spaces. 

    \item It can be parallelized reducing computation time. In our case study discussed in Sec. \ref{sec:parallel}, we observe that in order to achieve the same performance as the serial active algorithm, the parallelized version requires more simulations (about 50$\%$ in our examples), but this can still accomplish a speed up with respect to a uniform grid calibration similarly parallelized (up to a factor of 3 in our examples). 

    \item It can be used to improve models initially calibrated using a different strategy, say, with a uniform grid. In our example in Sec. \ref{sec:hybrid}, we observe an order of magnitude performance improvement when switching from uniform to active training. 

    \item It provides a quantitative, local and easy to interpret measure of uncertainty for the wave forms across parameter space. We moreover find that the average uncertainty over test set decreases as a power-law with the number of simulations used for training, and is significantly smaller for active training than for uniform or random grids. 

    \item By making the construction of the training set systematic, it reduces human bias in the construction of the approximants. While it is true that there is certain freedom in the choice of the acquisition function and other implementation details, the criteria for doing so can be succinctly expressed in an intuitive way. 

    \item The interpolation algorithm easily accommodates complex data patterns as it is based on GP regression, a non-parametric algorithm. In the simple case considered, we see a power-law decrease of the fit inaccuracy with number of simulations. We expect this feature to have the highest impact in cases which display intricate data patterns. Note that in order to fully exploit this feature, the acquisition function used should favour regions of higher parameter variation. We leave a careful study of this aspect for future work.

    \item  Our framework can be seamlessly integrated with different GW approximants. More generally, our approach can be deployed in conjunction with any model of temporal series. 

\end{itemize}

\section{Discussion}
\label{sec:discussion}

Only after the application of the proposed technique to more realistic scenarios we shall be able to properly evaluate its merits. With this in mind, let us discuss some potential limitations we identify in our study:

\begin{itemize}
    \item We assume that the model can perfectly describe the simulated data. We have made this choice since it allowed us to disentangle the effect of calibration from model improvement. However, in more realistic applications, there could be difficulties arising from model incompleteness or incorrectness, which are known to affect Bayesian adaptive experimental designs \cite{sloman2022characterizing}. 
    This limitation appears to have the lowest impact in the case of NR surrogates, since they are the most agnostic models, although they require larger training sets. 
    %

    \item  We use GP regression to fit the data for all choices of training sets. It is possible that for some specific models, using parametric fitting procedures (say, based on ratios of polynomials) could yield more accurate waveforms in some regions of parameter space. In particular, it might be that these simpler fits allow for extrapolation outside of the training region. However, we do not expect this to be a robust feature, since it is in particular contingent to the choice of specific model parameters.  

    \item Our computational cost and time estimates are based on the total number of simulations required by each training strategy. It is well-known, however, than some simulations are more costly than others (e.g. quasi-circulars with higher mass ratios). Since our main results involve relative costs, we expect that our conclusions should hold even when more precise cost estimates are taken into account. 

    \item We are assuming that the extraction of the model parameters from NR data is exact, since we read it off from the code output. In practice, this process involves some extra manipulations such as computing time derivatives of the signals, performing fits in the time domain, etc. Having available NR data of sufficient quality should reduce the impact of this limitation. 
    
\end{itemize}

Let us also consider some possible scenarios in which GPAL may not be most efficient when applied to GW modelling:
\begin{itemize} 

    \item  Given enough resources to parallelize the data collection of a large training set, our algorithm will eventually require more computational time than other non-adaptive approaches. This is an unlikely scenario given the high processing and memory requirements of the simulations. Moreover, as explained in Sec. \ref{sec:parallel}, this can be largely overcome by parallelizing the GPAL algorithm.

    \item  We have observed that using the GP variance as acquisition function tends to select points at the edges of the computational domain, which is typically where simulations are most challenging to carry out. In the case of large $q$, this incurs in higher computational cost, while large spins $|\chi| \sim 1$ present difficulties in the determination of the initial data. We thus recommend the use of this algorithm in regions of parameter space in which all simulations are feasible to carry out. Furthermore, our setting does not allow for extrapolation outside of the training domain. 
    
\end{itemize}

\section{Outlook}
\label{sec:outlook}

We identify several research questions which require further investigation:

\begin{itemize}

    \item Our results suggest that it is possible to efficiently improve existing models of GW by enlarging their calibration set of NR simulations as suggested by the Active Learning algorithm. It would be interesting to quantify this potential improvement for the available GW approximants for quasi-circular binary BH.  
    
    \item The case of BH dynamical captures (or highly eccentric binaries) \cite{East:2012xq, Rasskazov:2019gjw, Tagawa:2019osr} has been much less developed than that quasi-circular case, but recent progress has been made \cite{Nagar:2020xsk, Andrade:2023trh, Carullo:2023kvj}. 
    In this scenario, the two BHs undergo a series of close encounters before merger, resulting in a much richer phenomenology which involves a larger parameter space, and displays significant complexity in the model parameters. GPAL could be useful to guide parameter exploration in this and similar scenarios.    

    \item The notion of waveform uncertainty 
    has been argued to play an important role in parameter estimation \cite{Moore:2014pda, Moore:2015sza}. It would be interesting to extend these analysis to other models in which these notion was previously not incorporated, e.g. EOB approximants. 
    Moreover, it could serve as a statistical weight to build model ensembles or meta-models combining several approximants, see e.g. \cite{10.5555/1875158, ensemble_regression}.
\end{itemize}

We hope to address some of these open problems in the near future. 

\section*{Acknowledgments}

We are grateful to Patricio Reyes for collaboration during the early stages of the project. We thank Sebastiano Bernuzzi for comments on an earlier version of this manuscript, and Juan Calder\'on-Bustillo, Raimon Luna, Toni Font, Gregorio Carullo and Alessandro Nagar for insightful discussions. 
The work of TA is supported in part by the Ministry of Science and Innovation (EUR2020-112157, PID2021-125485NB-C22, 
CEX2019-000918-M funded by MCIN/AEI/10.13039/501100011033), and by 
AGAUR (SGR-2021-01069).
RG is supported by the Deutsche Forschungsgemeinschaft (DFG) under Grant No. 406116891 within the Research Training Group RTG 2522/1. RG also acknowledges funding from the National Science Foundation under Grant No. PHY-2020275.
\noindent \TEOB{} is publicly available at {\small {\url{https://bitbucket.org/eob_ihes/teobresums/}}} \\

\bibliography{refs_local.bib, refs.bib}

\appendix 

\section{EOB parameters}
\label{app:eob}

In the model \TEOB{}, the EOB parameters discussed in the main text $c_i = \{A_{\rm mrg}, \omega_{\rm mrg}, A_{\rm NQC}, \omega_{\rm NQC}, \dot A_{\rm NQC}, \dot \omega_{\rm NQC} \}$ are taken to be functions of the following form 
\begin{equation}
   Y = Y^{0} Y^{\rm orb}(\nu) Y^{\hat{S}}(X_{12}, \hat{S})
\end{equation}
for $Y= \{A_{\rm mrg}, \omega_{\rm mrg}, A_{\rm NQC}, \omega_{\rm NQC}\}$, and
\begin{equation}
   Z = Z^{\rm orb}(\nu) + Z^{\rm \hat{S}} (X_{12}, \hat{S})
\end{equation}
for $Z = \{ \dot A_{\rm NQC}, \dot \omega_{\rm NQC} \}$\footnote{In practice, the leading order multipolar behavior of the amplitude is factorized out before fitting, see App.~2 of \cite{Nagar:2020pcj}. Similarly, it was found that it is more convenient to fit $Z/(\nu \omega_{22}^{\rm NQC})$, rather than $Z$ directly}.
In the expressions above $X_{12} = (m_1 - m_2)/M$ and $\hat{S} = (S_1 + S_2)/M^2$, and we denote with a ``$0$'' superscript the test-particle limit, while the ``orb'' and ``$\hat{S}$'' superscripts indicate the orbital and spin part respectively.
We further have that
\begin{align}
    Y^{\rm orb} &= 1 + a_1^{Y}\nu + a_2^{Y} \nu^2 \, , \\
    Y^{\rm \hat{S}} &= \frac{1 + b_1^{Y} \hat{S} + b_{2}^{Y} \hat{S}^2}{1 + b_3^{Y} \hat{S}} \, , \\
    b_{i}^{Y} &= \frac{c_{i,0}^{Y} + c_{i,1}^{Y}X_{12}}{1 + c_{i,2}^{Y}X_{12}} \, , \\
    Z^{\rm orb} &= 1 + a_1^{Z}\nu + a_2^{Z} \nu^2 \, , \\
    Z^{\rm \hat{S}} &= (b_1^Z + c_1^Z X_{12})\hat{S} + (b_2^Z + c_2^Z X_{12})\hat{S}^2 \, .
\end{align}
The set of quantities fit to NR are therefore the $\{ a^{Y}, c^{Y}, a^{Z}, b^{Z}, c^{Z}\}$ coefficients, where pedices have been dropped for brevity.

\section{GP Regression}
\label{app:GP}




In this appendix we collect some useful information about GP regression closely following \cite{10.7551/mitpress/3206.001.0001}. 
A GP is a set of random variables, any of which have a joint Gaussian (multivariate, normal) distribution. 
More concretely, if we consider a set of inputs $\{X \}$, we assume that a function $f$ takes values $f(X)$ which are normally distributed with some mean $m(X)$ (which for notational simplicity we take to be $0$) and a covariance function (also known as kernel) $K(X, X)$ which is symmetric and positive definite.
\begin{equation}
    f \sim \mathcal{N}(0, K(X, X))
\end{equation}
\noindent where $\mathcal{N}$ is the normal distribution. 

Similarly, for unobserved points $X_*$ (say, in the test set), the values $f_* := f(X_*)$ are distributed normally with the same kernel. 
The joint distribution of the observed (training) and unobserved (test) points is given by the prior
\begin{equation}
    \begin{bmatrix}
    f \\ f_* 
    \end{bmatrix} 
    \sim 
    \mathcal{N} \left(0, 
    \begin{bmatrix}
        K & K_* \\ K_* & K_{**}
    \end{bmatrix}
    \right)
\end{equation}
\noindent where $K = K(X,X)$, $K_* = K(X_*, X)$, $K_{**} = K(X_*,X_*)$.
To get the posterior distribution used for regression, we condition over the observed points, which after some manipulations, yields
\begin{align}
\label{eq:app_GP_post}
    f_* | X_*, X, f \sim \mathcal{N} (K_* K^{-1} f, 
    K_{**} - K_*K^{-1}K_*)
\end{align}
The predictions and variance of the GP regression are the mean 
and variance of the distribution in \eqref{eq:app_GP_post}. Note that the variance only depends on the training and test sets, but not on the function being sampled. 

In our numerical experiments, we have used the kernel denoted Matern52 with Auto Relevance Determination (ARD), which is given by the expression 
\begin{equation}
     k(r) = \sigma^2 (1 + \sqrt{5} r + \frac53 r^2) \exp(- \sqrt{5} r)
\end{equation}
\noindent with 
\begin{equation}
    r^2(x, x') = \sum_{q = 1}^Q \frac{(x_q - x_q')^2}{\ell_q^2}
\end{equation}
\noindent where the $\ell_q$ are relative length scales automatically determined for each direction and $Q$ the number of dimensions in parameter space. We fix $\sigma = 1$ throughout. 
Note that the correlation decays exponentially with the distance between inputs, which ensures the smoothness of the output functions. 
We have used the implementation provided by the Python library GPy \cite{GPy}.

\section{Performance and uncertainty metrics}
\label{app:metrics}

Here we provide details regarding the performance and uncertainty metrics used in the main text. 
We begin with the faithfulness, or match, which is defined as
\begin{equation}
    {\cal F} = \frac{\langle h_1 , h_2\rangle}{\sqrt{\langle h_1 , h_1 \rangle \langle h_2 , h_2\rangle}}
\end{equation}
\noindent here the inner product product between two waveforms $h_1$, $h_2$ is given by
\begin{equation}
    \langle h_1 , h_2\rangle = 4 \Re \int \frac{\hat h_1(f) \hat h_2^*(f)}{S_n(f)}
\end{equation}
\noindent where $\hat h_{1,2}(f)$ are the Fourier transforms of the time domain signals. Throughout this paper we consider a uniform PSD $S_n(f) = 1$, but more generally this is where the noise curve of a particular GW detector enters the calculation. The unfiathfulness is then given by $\overline{{\cal F}} = 1 - {\cal F}$. 

We also consider the amplitude and phase difference at merger. To obtain these, we decompose the signal in amplitude and phase as
\begin{align}
    h_{gt}(X, t) & = A_{gt}(X, t) e^{- i \phi_{gt}(X, t) } \\
    \tilde h(X, t) & = \tilde A(X, t) e^{- i \tilde \phi(X, t) }
\end{align}
\noindent and compute the difference at merger, defined as the amplitude peak of each signal, 
\begin{align}
    \delta A(X) &= (A_{gt}(X, t_{peak}) - \tilde A(X, t_{peak}))/A_{gt}(X, t_{peak}) \\
    \delta \phi(X) &= \phi_{gt}(X, t_{peak}) - \tilde \phi(X, t_{peak})
\end{align}

The most uncertain signal for a given variance is defined by first computing the family of waveforms
\begin{equation}
\label{eq:h_alpha}
    \tilde h^{(\alpha)}(X, t) = h_{EOB}\left(\tilde c_i(X) + \frac{1}{2}\Delta^{(\alpha)}_i \sigma(X); t \right)
\end{equation}
\noindent where $\Delta^{(\alpha)}_i$ account for all possible sign combinations of internal parameters, e.g $\{+,+,+,+,+,+\}$, $\{-,+,+,+,+,+\}$, etc. For the case considered here having $6$ internal parameters, we need to compute $2^6 = 64$ different waveforms. Note that this captures the notion of directional derivative of the waveform with respect to the $c_{i}$. We then select the most uncertain signal as the representative $\tilde h^{(\alpha_0)}(X, t)$ that maximises the norm $|\tilde h^{(\alpha)}(X, t) - \tilde h^(X, t)|$, given by
\begin{equation}
\label{eq:norm}
    |h_1(t) - h_2(t)| = \int dt | h_1(t) - h_2(t) |^2
\end{equation}
\noindent within this set.

In order to quantify the accuracy of the fit, we  also introduce the Euclidean norm of the difference between the ground truth values of the parameters and the fitted ones, 
\begin{equation}
\label{eq:delta_c}
    \delta c(X) = ||c_i(X) - \tilde c_i(X)||
\end{equation}

\section{Additional numerical results}
\label{app:more}

\subsection{Alternative test sets}

In the main text we reported the results obtained for a test set consisting of the reciprocal lattice associated to the uniform training set for given $N_{{\rm train}}$, with the boundary points added, see Fig. \ref{fig:training_test_coords}. In this appendix we explore other possibilities for test sets, for the same training strategies (active, uniform, random). 
As in the main text, we use as reference values for the performance metrics $\overline{{\cal F}} = 0.01$, $\delta A = 0.1$, $\delta \phi = 0.06$. 

\begin{figure}[] 
\centering
\includegraphics[width=0.4\textwidth]{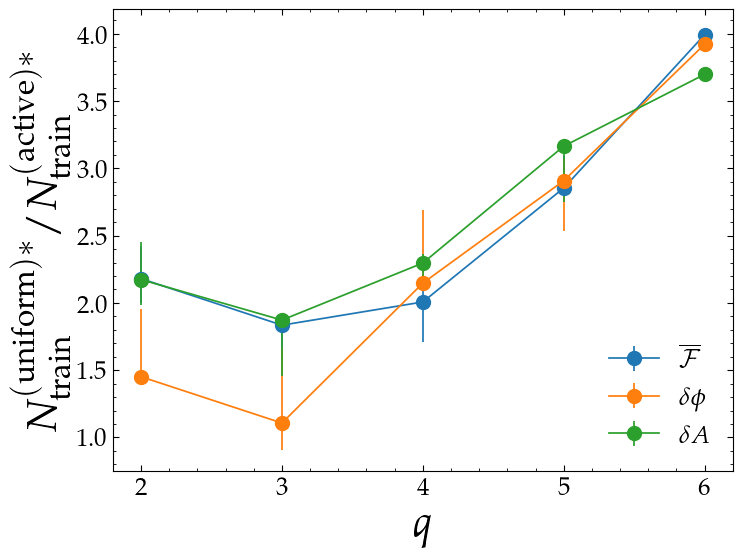}
\includegraphics[width=0.4\textwidth]{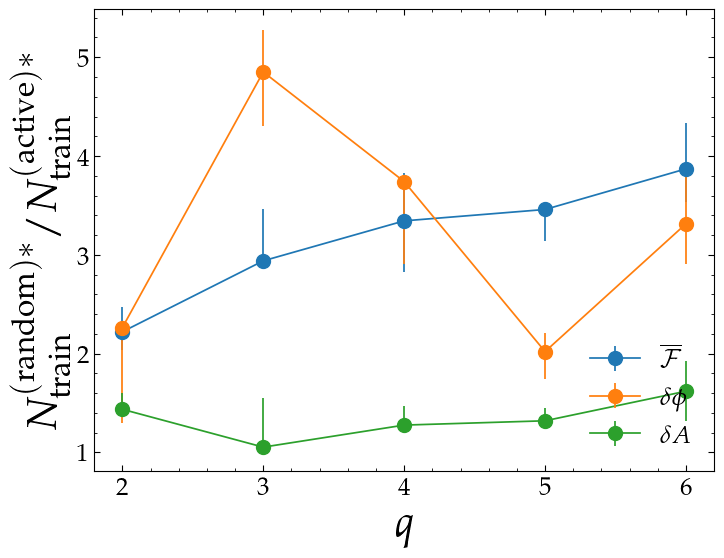}
\caption{Average performance gain of active training for test sets constructed as the reciprocal lattice used for uniform training (top) and for fully random test sets with $N_{{rm test}} = 200 $ (bottom). In the latter case, random training outperforms uniform training, so we report the  performance gain with respect to that training strategy.} 
\label{fig:app_alt_test}
\end{figure}

We first consider test sets such as the one discussed in the main text, but with the boundary points removed. This is motivated by the fact that boundary points are the most challenging to model (see Appendix \ref{app:local}), which could lead to a misrepresentation of the performance if one is interested in ``typical'' points. 
We show the relative performance gain for the test set in the main text with the boundary points removed in Fig. \ref{fig:app_alt_test}, to be compared with Fig. \ref{fig:performance}. Besides the reduction of the error bars in the test set without boundary points, we observe essentially the same behaviour, i.e. the performance gain grows with $q_{{\rm max}}$ reaching a factor of 4x for the largest value $q_{{\rm max}}= 6$.

We have also considered test sets chosen at random with uniform distribution across parameter space. In such cases, we typically observe that random training outperforms uniform training, although active training continues to be superior. We show in Fig. \ref{fig:app_alt_test} the average performance gain of active training over random training. In the case of the unfaithfulness, we observe a similar behaviour to the reciprocal lattice test, with a monotonic growth of performance reaching nearly 4x. For the phase difference at merger, the performance gain is not monotonic and has larger error bars, which could be explained by the randomness of the test sets. For the amplitude difference at merger, we find a very mild performance gain slightly above 50$\%$. 

\subsection{Alternative training sets}

As discussed in Sec. \ref{sec:calibration}, the active training strategy tends to select points near the boundary of the computational domain. This suggests that it might be possible to improve the performance of greedy approaches by replacing a uniform grid for a Chebyshev one, which is denser near the edges\footnote{We thank Sebastiano Bernuzzi for suggesting this numerical experiment.}. For the interval $(a, b)$ the Chebyshev nodes are given by 
\begin{equation}
    x_k = \frac{1}{2}[ (a+b) + (a-b) \cos \langle( k \pi/n  \rangle) ] \qquad k = 0 .. n
\end{equation}

\begin{figure}[h] 
\centering
\includegraphics[width=0.4\textwidth]{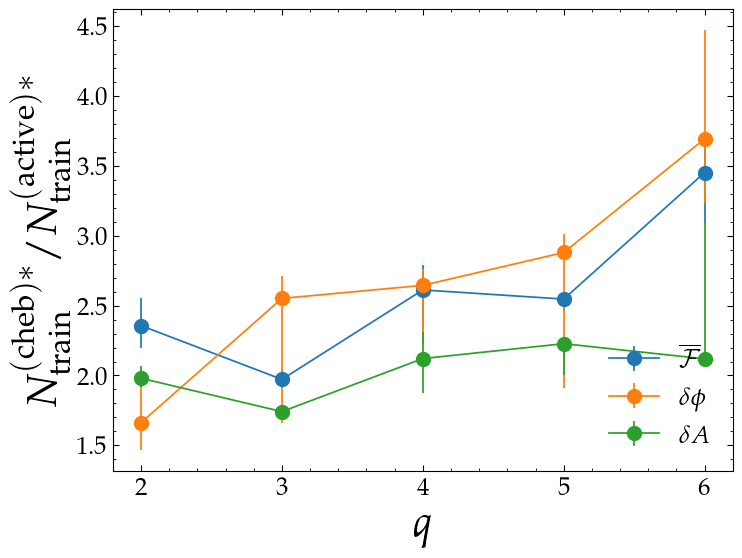}
\caption{Average performance gain of active training over Chebyshev training as a function of $q_{{\rm max}}$. All test sets are random grids with the same number of points as the training sets.} 
\label{fig:perf_gain_cheb}
\end{figure}

\begin{figure*}[] 
\centering
\includegraphics[width=0.8\textwidth]{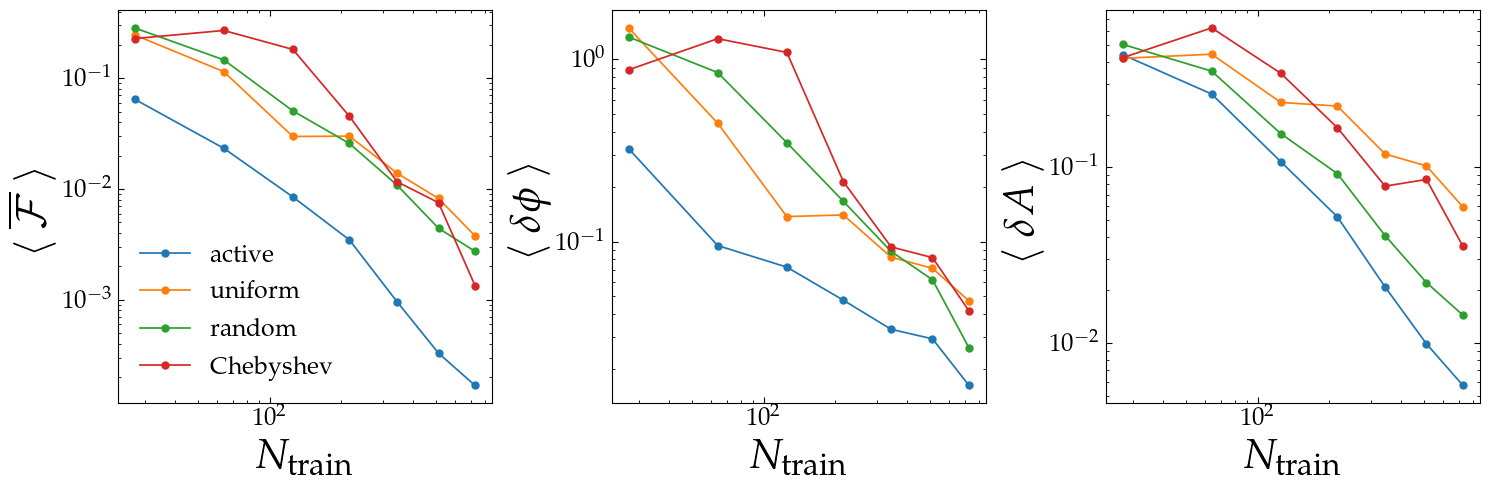}
\caption{Dependence of the average performance metrics with number of training points, for active, uniform, random and Chebyshev training with $q_{{\rm max}} = 6$. The test sets are random grids the same number of points as the training sets. All plots are in log-log scale.} 
\label{fig:metrics_vs_Ntrain_cheb}
\includegraphics[width=0.8\textwidth]{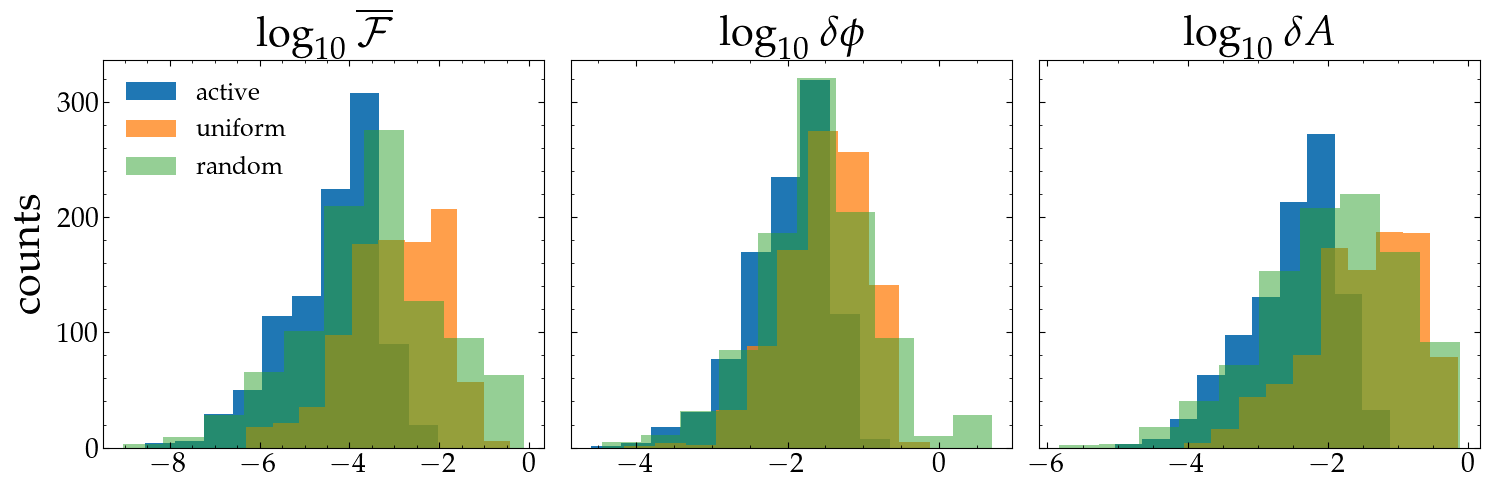}
\includegraphics[width=0.8\textwidth]{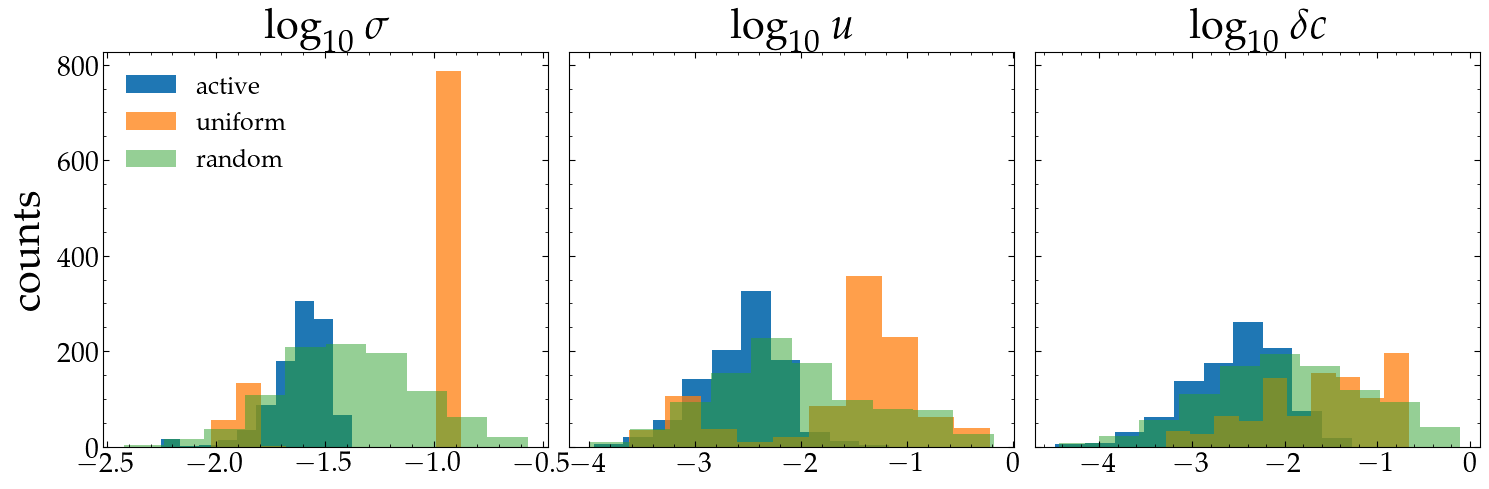}
\caption{Histograms of the performance (top) and uncertainty metrics (bottom) for $N_{{\rm train}} = 729$ with $q_{{\rm max}} = 6$.} 
\label{fig:app_hist_metrics}
\end{figure*}

In addition to the active, uniform and random training strategies discussed in detail in the main text, we consider numerical experiments with a (passive) Chebyshev training consisting of a 3d grid with Chebyshev nodes along each direction, having the same number of training points as all other approaches. 
Using the reciprocal lattice of the uniform training as test set gives a biased evaluation of the different algorithms since the Chebyshev points tend to be close to the reciprocal lattice. We therefore consider fully random test sets, with the same number of points as the corresponding training sets. 
We show our results for the average performance gain of active training over Chebyshev training in Fig. \ref{fig:perf_gain_cheb}, and the average performance as a function of training points for $q_{{\rm max}} = 6$ in Fig. \ref{fig:metrics_vs_Ntrain_cheb}. We observe that uniform, random, and Chebyshev training give similar results, all being superseded by active training. The maximum cost reduction obtained by using active training is again roughly a factor of 4 for the mismatch, and slightly lower for other metrics. 

\section{Local results}
\label{app:local}
\begin{figure*}[] 
\centering
\includegraphics[width=0.4\textwidth]{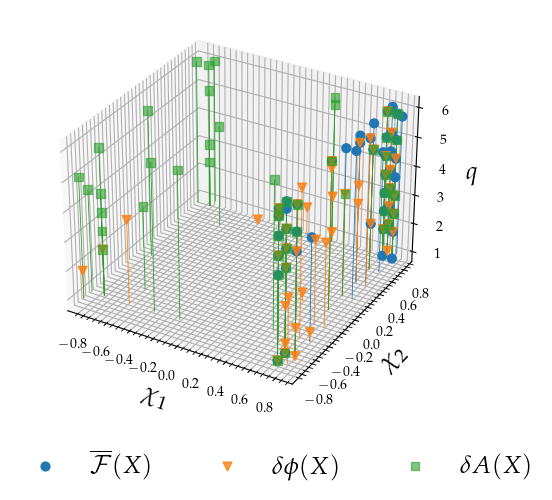}
\includegraphics[width=0.4\textwidth]{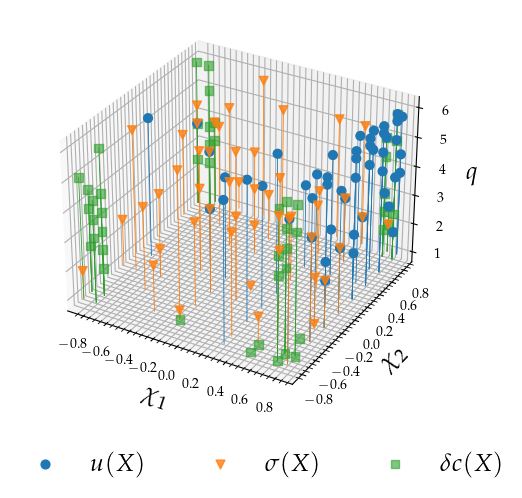}
\includegraphics[width=0.4\textwidth]{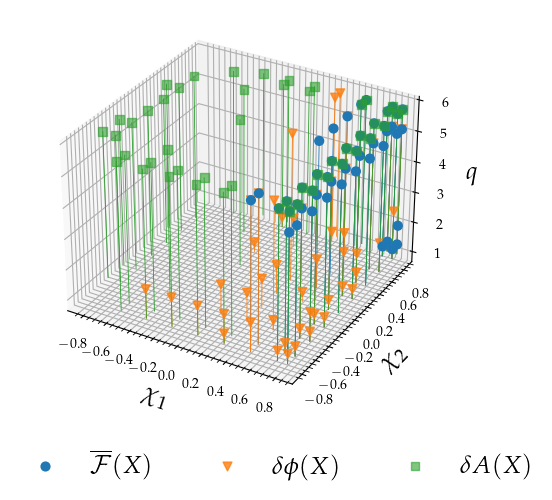}
\includegraphics[width=0.4\textwidth]{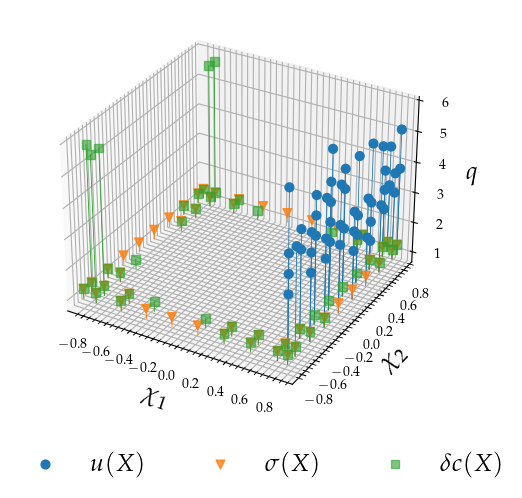}
\includegraphics[width=0.4\textwidth]{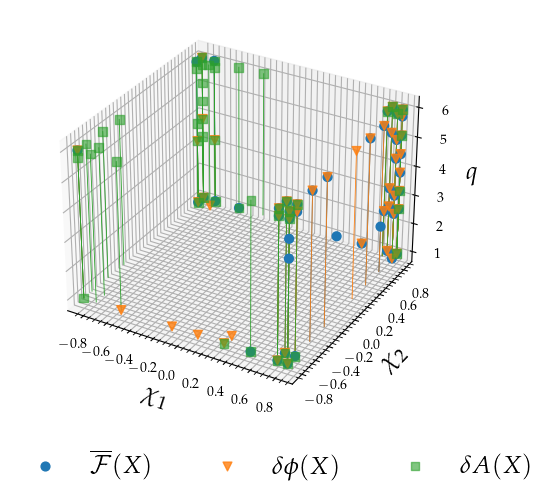}
\includegraphics[width=0.4\textwidth]{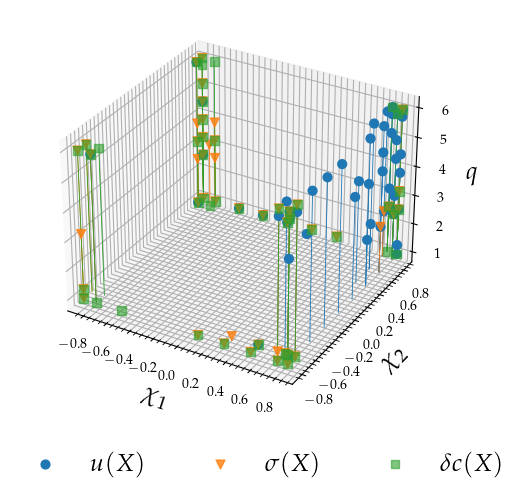}
\caption{Three-dimensional local dependence of the performance (left column) and uncertainty (right column) metrics for active (top row), uniform (center row) and random (bottom row) training with $N_{{\rm train}} = 729$. We show the 50 worst points in each case.} 
\label{fig:worst_local}
\end{figure*}

In the main text we focused on the averages of performance and uncertainty metrics. In this appendix we complement these results with the statistical and spatial distributions in parameter space.  
First, we show in Fig. \ref{fig:app_hist_metrics} histograms of all metrics extracted for $N_{{\rm train}} = 729$. We Note that not only the averages are always smaller for active training, but the entire distributions show the superior performance of this protocol over uniform and random grid training. 
In Fig. \ref{fig:worst_local} we display a visualization of the local dependence of all metrics as functions of the input parameters. We observe that, with the exception of the variance in active training, the most challenging cases tend to accumulate near the boundaries of the computational domain.


\end{document}